\pdfoutput=1 
\documentclass[final,1p]{elsarticle}	

\usepackage{graphicx}
\usepackage{amssymb}
\usepackage{bm}
\usepackage{amsmath,amsfonts,latexsym,color}

\usepackage{braket} 

\usepackage{comment}

\usepackage{bm} 

\newcommand*{\br}{\mathbf{r}}

\newcommand*{\bk}{\mathbf{k}}

\newcommand*{\cI}{{\cal I}}

\newcommand*{\phop}{\phi^{\vphantom{\dagger}}}
\newcommand*{\phdop}{\phi^\dagger}
\newcommand*{\aop}{a^{\vphantom{\dagger}}}
\newcommand*{\adop}{a^\dagger}

\newcommand*{\cop}{c^{\vphantom{\dagger}}}
\newcommand*{\cdop}{c^\dagger}

\newcommand*{\pop}{p^{\vphantom{\dagger}}}
\newcommand*{\pdop}{p^\dagger}
\newcommand*{\hop}{h^{\vphantom{\dagger}}}
\newcommand*{\hdop}{h^\dagger}

\newcommand*{\phdagger}{\vphantom{\dagger}}

\begin{document}

\begin{frontmatter}
	
\title{Potential insights into non-equilibrium behavior from atomic physics}

\author{Austen Lamacraft} 
\address{Department of Physics, University of Virginia,
Charlottesville, VA 22904-4714 USA}

\author{Joel Moore}
\address{Department of Physics, University of California,
366 Le Conte Hall,
Berkeley, CA 94720-7300 USA}

\date{\today}

\begin{abstract}	

This chapter seeks to outline a few basic problems in quantum statistical physics where recent experimental advances from the atomic physics community offer the hope of dramatic progress.  The focus is on nonequilibrium situations where the powerful concepts and methods of equilibrium statistical physics and ``linear response'' theory (for small deviations from equilibrium) are not applicable.  The problems discussed here are chosen in part because they have a high degree of ``universality'' or generality across different microscopic situations, as the major challenge in nonequilibrium statistical physics, both quantum and classical, has been to find principles as general as the basic principles of equilibrium statistical physics or linear response.

\end{abstract}

\end{frontmatter}

\section{Introduction}
\label{sec:introduction}

The initial engagement of the condensed matter community with ultracold atomic physics during the first half of the past decade was due to the apparent \emph{similarity} between the newly-created atomic systems and familiar (electronic) counterparts in the solid state. The two most notable examples of this trend were the observation of the superfluid to Mott insulator transition in an optical lattice \cite{greiner:2002}, and the creation of superfluid states of paired fermions, the neutral analogs of s-wave superconductors \cite{regal:2004}. 

One feature of the latter half of the decade was a growing appreciation for the \emph{differences} between these two settings for quantum many body physics. This chapter concerns one of the most prominent, intriguing, and conceptually far-reaching of these: the out of equilibrium nature of many experiments in ultracold physics. Mechanisms of equilibration in solid state systems are typically fast -- on the picosecond timescale\footnote{Glassy systems are of course a notable exception.} -- with the resulting equilibrium states essentially immortal. In contrast, the lower temperatures and energy scales of ultracold gases means that the corresponding processes can be observed at the millisecond level, and non equilibrium physics is a fact of life during the few second lifespan of most ultracold systems.

Furthermore, there is good reason to believe that the \emph{mechanisms} of equilibration in the ultracold domain are distinct from those that are generically important for electrons in the solid state. There are no phonons that can transfer energy into the lattice, (typically) no impurities to allow momentum to dissipate, and no spin-orbit interaction to mediate spin relaxation. We find ourselves therefore in an enviable (if unfamiliar) position: the simple Hamiltonians that we write for the degrees of freedom of interest are, to a good approximation, all there is. To take a simple example, an applied Zeeman field will typically polarize the spins of an electron system, while in an atomic system it leads only to Larmor precession (the quadratic Zeeman effect can elicit interesting effects, however).

The study of equilibration in \emph{isolated} systems (though this term should be used carefully, see below) described by strictly Hamiltonian dynamics is of course as old as statistical mechanics itself, dating back at least to Boltzmann. It is the possibility of studying non equilibrium \emph{quantum} phenomena in a very simple setting that has led to a resurgence of interest in this problem of fundamental physics \cite{Cazalilla:2010,Polkovnikov:2010}. Additionally, the way in which a system comes to equilibrium is believed to be sensitive to a number of factors that are already the focus of many experimental investigations in ultracold systems, namely \emph{dimensionality}, \emph{disorder}, \emph{integrability} of the underlying dynamics, and of course the \emph{initial conditions}. 


In trying to draw together some of these threads for this chapter, we are faced with the familiar problem: equilibrium systems are all alike, while every non-equilibrium system is out of equilibrium in its own way\footnote{With apologies to Tolstoy.}. Nevertheless, most phenomena of current interest can be associated with one or more of the above four aspects. Let us take as an example an experiment that has acquired an iconic status as an illustration of out of equilibrium behavior in an atomic gas. In the `quantum Newton's cradle' of Kinoshita \emph{et al.} \cite{Kinoshita:2006}, the evolution of the momentum distribution of arrays of harmonically confined 1d Bose gases was studied over many periods of oscillation, after initially splitting the gas into two counterpropagating clouds. Even though each atom undergoes many thousands of collisions during this time, the momentum distributions do not relax to the equilibrium distribution (determined to be Gaussian for thermal clounds of the same rms momentum). This is to be contrasted with thermalization in a three-dimensional gas, which occurs after a very small number of collisions.

A plausible argument for this behavior is easily made if one assumes the applicability of the Boltzmann equation in its naive form (which is by no means clear), with a collision integral describing two-particle collisions. In a gas of particles of equal mass, such collisions do not lead to a change in the distribution function, as particles either retain or exchange their momenta. Within the Boltzmann picture, a change in the distribution function requires at least three-body collisions to be accounted for, as was done in a number of recent works \cite{Mazets:2008,Tan:2010}. According to the discussion of Ref.~\cite{Mazets:2010}, once the suppression of three-body scattering due to interparticle repulsion is accounted for, the resulting damping rates are consistent with the long times observed by Kinoshita \emph{et al.}.

The example of the quantum Newton's cradle vividly illustrates the role played by reduced dimensionality. Additionally, the experimental conditions closely approximate one of the best-known integrable many-body systems: the 1d Bose gas with $\delta$-function interaction\footnote{Also known as the Lieb--Liniger gas \cite{lieb1963}.}. In this integrable system, three-body collision are absent, or more correctly factorize into successive two-body collisions. The three-body collisions that will eventually cause the system to equilibrate arise because the 1d model is only an approximation to experimental reality. The potential confining atoms to a tube naturally has higher modes of transverse excitation, and it is virtual processes involving these modes that gives rise to the three-body interactions \cite{Mazets:2008}. In Section~\ref{sec:integrability} we will see another example of thermalization due to the breaking of integrability. 


Equilibration is also often deemed to be particularly precarious in \emph{disordered} systems, which appear at first to be the polar opposites of integrable systems. The phenomenon of localization due to quantum mechanical interference has been studied constantly since Anderson's work \cite{anderson:1958} more than fifty years ago. Recently it was suggested \cite{basko:2006} that an \emph{isolated} disordered interacting system can undergo a finite temperature localization transition. In the low temperature (localized) phase the system is unable to come to internal equilibrium. This fascinating idea would seem to have a natural home in ultracold systems and will be discussed in Section~\ref{sec:loc}.

We begin our survey with the last of the themes identified above: the choice of initial conditions. A natural way to initiate out of equilibrium dynamics is to abruptly change some system parameter. In the classical description of phase ordering one speaks of a \emph{quench} if the variation of this parameter would cause the equilibrium system to pass through a phase transition. By analogy, abruptly crossing a quantum phase transition has become known as a \emph{quantum quench}, an alliterative coinage apparently due to Calabrese and Cardy \cite{Calabrese:2006}, though recent papers use the term to refer to any abrupt change in the system. Section~\ref{sec:quench} gives an introduction to some of the general theoretical issues with reference to a simple model, the `$\lambda \phi^{4}$' field theory.



As indicated above, this survey is necessarily incomplete and many interesting aspects of non-equilibrium behavior are missing. For example, we will have nothing to say about non-equilibrium steady states, which seem less relevant to ultracold physics than to (say) mesoscopic physics, where conditions of constant drive can be sustained for long periods.  We will only have a few brief comments about connections between quantum non-equilibrium physics and quantum information concepts such as entanglement.   It is hoped that readers will find the chapter a useful effort to motivate and explain some key current questions in non-equilibrium physics that are ripe for investigation.

\section{Quantum Quenches}\label{sec:quench}

\subsection{Introduction} 
\label{sub:introduction}



The study of \emph{phase ordering kinetics} has an extremely long history in classical statistical physics \cite{Bray:2002}. The main theoretical problem is to understand how a new phase appears after an abrupt change in system parameters. For a symmetry-breaking transition, this entails an understanding of the growth of the order parameter, and the dynamics of topological defects formed in the process. It is natural to ask what changes when we cross a quantum -- rather than classical -- phase transition, and more generally how quantum mechanics changes things.

We would like to argue that for a symmetry-breaking transition, it is only in the early stages of phase ordering that the difference will be acute\footnote{Given the relatively short lifetimes of ultracold systems, it is also likely to be the only part of the story accessible to experiment for the foreseeable future.}. The reason is that the initial dynamics establishes a local fluctuating order parameter distribution that soon becomes macroscopic in character and leaves the quantum domain. The subsequent evolution of this distribution (often called \emph{coarsening} in the literature) will then be described by classical equations of motion\footnote{We hope the meaning of \emph{classical} is clear: superfluid hydrodynamics is a classical theory though superfluidity has a quantum origin.}. Thus the following discussion will focus on the quantum dynamics of the early stages of phase ordering and the emergence of the classical description within a simple model. We should point out that this issue was discussed many years ago in a cosmological context \cite{Guth:1985}, predating the more recent work motivated by the experiment of Ref.~\cite{Sadler2006} \cite{Lamacraft:2007,Uhlmann:2007,Saito:2007,Mias:2008}.

\subsection{A model system} 
\label{sub:a_model_system}

Let try to describe the effect of a quench in a system described by the `$\lambda \phi^{4}$' Hamiltonian
\begin{equation}
	\label{quenches_model}
	H = \frac{1}{2}\int d\br \left[|\nabla\Phi|^{2}+|\Pi|^{2}+m^{2}|\Phi|^{2}+\lambda|\Phi|^{4} \right],
\end{equation}
where $\Phi(\br)$ and $\Pi(\br)$ are complex canonically conjugate fields $\left[\Phi(\br),\Pi(\br')\right]=i\delta(\br-\br')$. This model describes a quantum phase transition between a disordered phase ($\langle\Phi\rangle =0$) at positive $m^{2}$ to an ordered one ($\langle\Phi\rangle\neq 0$) at negative values. For simplicity, we imagine that initially the system is in the ground state corresponding to $m^{2}=m_{i}^{2}>0$. 

Ultracold physics provides two examples of systems with transitions described by the model Eq.~\eqref{quenches_model}. The first is the superfluid insulator transition in the Bose--Hubbard model \cite{fisher:1989,greiner:2002}, where a non-zero order parameter $\langle\Phi\rangle$ corresponds to the condensate wavefunction in the superfluid phase, and the transition may be tuned by varying the depth of the optical lattice that confines the bosons. The second example is provided by a condensate of spin-1 bosons with ferromagnetic spin-spin interactions. If the system has zero total magnetization with respect to the axis defined by the applied magnetic field, only the quadratic part of the Zeeman energy is effective and leads, at large fields, to a condensate in the $m=0$ state. As the field is lowered, the ground state of the condensate makes a transition to a state with nonzero transverse magnetization \cite{stenger:1998,murata:2007,Lamacraft:2007}. If we consider only small changes of the magnetic field around the critical value, this problem maps onto the `$\lambda \phi^{4}$' theory where $\langle\Phi\rangle$ gives the (amplitude and phase of the) transverse magnetization \cite{Lamacraft:2007}. The groundbreaking experiment described in Ref.~\cite{Sadler2006} explored the dynamics following a quench in this system.

Thus both realizations have a natural control parameter corresponding to $m^{2}$, that can be quenched from positive to negative values. The `$\lambda \phi^{4}$' theory is of course one of the first examples of a quantum field theory that one meets, and a prototype for studies of spontaneous symmetry breaking. Therefore it is not surprising that the \emph{dynamics} of spontaneous symmetry breaking in this theory initiated by a quantum quench in $m^{2}$ has been rather extensively studied in the particle physics and cosmology literature (see for example Ref.~\cite{cooper:1997} and references therein), where the principal motivation is to develop intuition for phase transitions in the early universe. 

There is a hierarchy of approximations of increasing complexity that have been applied to this and related problems, reflecting the difficulty of extending the description of the behavior of the system to longer and longer times. 
As discussed above we will focus dynamics immediately following a quench, though we close this section by discussing the theoretical approaches that have been used to understand the late-time behavior.


\subsection{Instability at the Gaussian level: emergence of classicality for one mode} 
\label{sub:instability_at_the_gaussian_level}


The very simplest approximation involves discarding completely the quartic term in Eq.~\eqref{quenches_model}, leaving a Gaussian model that may be solved exactly. Understanding the dynamics after a quantum quench in this model is still a useful exercise, however. Since the Hamiltonian separates into a sum of harmonic oscillator Hamiltonians for modes of each wavevector, we start by considering a single oscillator.
\begin{equation}
	\label{quenches_single_SHO}
	H = \frac{1}{2}p^{2}+\frac{1}{2}\omega^{2}x^{2}
\end{equation}
Let us suppose that at time $t=0$ the frequency is abruptly changes from a real to an imaginary value $\omega_{i}^{2}>0\to \omega_{f}^{2}<0$. The subsequent evolution of the operators $x(t)$ and $p(t)$ in the Heisenberg picture coincides with the classical solution
\begin{gather}
	\label{quenches_sho_sol}
	x(t)=x(0)\coth\left(|\omega_{f}|t\right)+\frac{p(0)}{|\omega_{f}|}\sinh\left(|\omega_{f}|t\right)\xrightarrow[t\to \infty]{}\frac{e^{|\omega_{f}|t}}{2}\left(x(0)+\frac{p(0)}{|\omega_{f}|}\right)\\
	p(t)=p(0)\coth\left(|\omega_{f}|t\right)+x(0)|\omega_{f}|\sinh\left(|\omega_{f}|t\right)\xrightarrow[t\to \infty]{}\frac{e^{|\omega_{f}|t}}{2}\left(p(0)+|\omega_{f}|x(0)\right).
\end{gather}
%

Condensed matter physicists would be inclined to call the resulting time evolution the unfolding of a Bogoliubov transformation, while quantum opticians know this operation as \emph{squeezing}.

This solution may be used to evaluate any desired correlation function, using the initial correlation functions at time $t=0$ corresponding to the ground state with $\omega^{2}=\omega_{i}^{2}$. For example
\begin{align}
	\label{quenches_onemode}
	\langle x(t)^{2}\rangle &= \coth^{2}\left(|\omega_{f}|t\right)\langle x(0)^{2}\rangle+\frac{\langle p(0)^{2}\rangle}{|\omega_{f}|^{2}}\sinh^{2}\left(|\omega_{f}t|\right)\\ 
	& = \frac{1}{2\omega_{i}}\coth^{2}\left(|\omega_{f}|t\right)+\frac{\omega_{i}}{2|\omega_{f}|^{2}}\sinh^{2}\left(|\omega_{f}t|\right)\\
	&\to \frac{1}{8}\left[\frac{1}{\omega_{i}}+\frac{\omega_{i}}{|\omega_{f}|^{2}} \right]e^{2|\omega_{f}|t},\, \text{as } t\to\infty.
\end{align}
The exponential growth is exactly what would be expected from the instability of the classical system, though of course $\langle x(t)\rangle=0$ by symmetry. Quantum mechanics brings the novel feature that the instability is effective even in the absence of noise or some other perturbation to push the system away from the unstable maximum: the zero point motion in the initial state is enough. 

The quantum origin of the initial `seed' notwithstanding, the description of the system becomes increasingly classical as time progresses. To understand why this is so, notice that the $t\to\infty$ limits of $x(t)$ and $p(t)$ given in Eq.~\eqref{quenches_sho_sol} are commuting. Now of course, the evolution is unitary, so $\left[x(t),p(t)\right]=i$, but this lack of commutativity is increasingly irrelevant as time progresses, so that in the long time limit one can take $p(t)=|\omega_{f}|^{2}x(t)$, with $x(t)$ having a Gaussian distribution with the above variance.

\begin{figure}
	\centering
		\includegraphics[width=0.75\columnwidth]{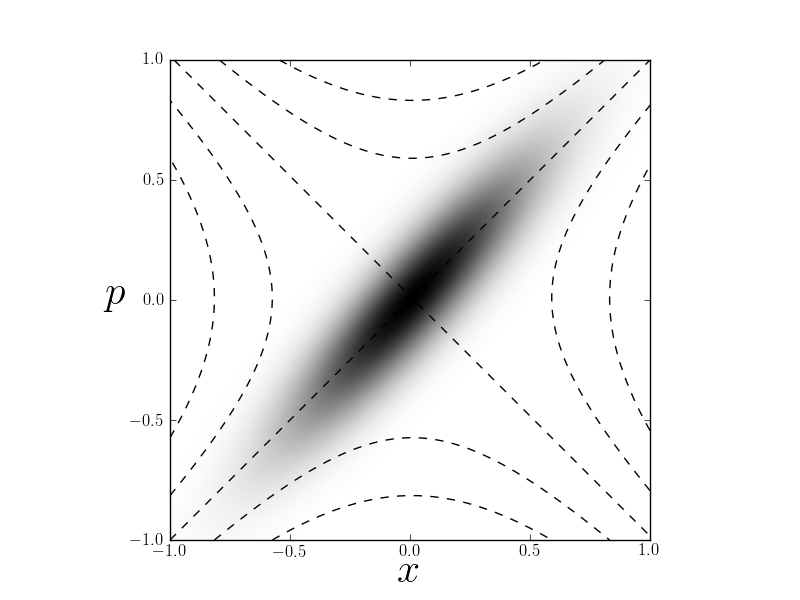}
	\caption{The Wigner distribution in Eq.~\eqref{quenches_wigner} with $\omega_{i}t=0.7$, $|\omega_{f}|=\omega_{i}$. The dashed lines indicate the contours of the Hamiltonian $H=\frac{1}{2}p^{2}+\frac{1}{2}\omega_{f}^{2}x^{2}$. The initial distribution is axially symmetric but is `squeezed' along the classical trajectory with the passage of time.}
	\label{fig:wigner}
\end{figure}

It's also useful to consider the Schr\"odinger picture counterpart of the evolution in Eq.~\eqref{quenches_sho_sol}. The initial Gaussian is annihilated by $x(0)+ip(0)/\omega_{i}$, so the state at time $t$ is likewise annihilated by $x(t)+ip(t)/\omega_{i}$, corresponding to the wavefunction
\begin{equation}
	\label{quenches_psisol}
	\psi(x,t)=A(t)\exp\left(-\frac{x^{2}|\omega_{f}|}{2}\tan(\theta+i|\omega_{f}|t) \right)\xrightarrow[t\to \infty]{} A(t)\exp\left(-\frac{ix^{2}|\omega_{f}|}{2}\left[1-2e^{i\theta}e^{-2|\omega_{f}|t}\right] \right)
\end{equation}
where $\tan\theta=\omega_{i}/|\omega_{f}|$, and $A(t)$ is a normalizing factor. The Wigner distribution of the associated density matrix $\rho(x_{1},x_{2},t)=\psi^{*}(x_{1},t)\psi(x_{2},t)$ is
\begin{align}
	\label{quenches_wigner}
\begin{split}
	f(p,x,t)&=\int dx_{12}\, \psi^{*}(x+x_{12}/2,t)\psi(x-x_{12}/2,t) e^{-ipx_{12}}\\
	&\propto\exp\left(-\alpha(t)x^{2}-\frac{\left[p-\beta(t)x\right]^{2}}{\alpha(t)}\right),\text{ as }t\to\infty
\end{split}
\end{align}
with $\alpha(t)=2|\omega_{f}|\sin 2\theta e^{-2|\omega_{f}|t}$, $\beta(t)=|\omega_{f}|\left(1-2\cos 2\theta e^{-2|\omega_{f}|t}\right)$. The form of this distribution is illustrated in Fig.~\ref{fig:wigner}. Because of the decay of $\alpha(t)$, the width of the $x$ distribution is increasing exponentially, while at long times the $p$ distribution is effectively $\delta(p-|\omega_{f}|x)$. In other words, it is nonzero only on the phase space trajectory $p=|\omega_{f}|^{2}x$ of a classical particle that starts at $p=x=0$.

The exponential growth in the width of the $x$ distribution also reflects the classical trajectory of the particle, so that the probabilistic character may be attributed to the initial state by writing
\[x(t)=x_{0}e^{|\omega_{f}|t}\]
where $x_{0}$ is a classical Gaussian random variable with variance $\langle x_{0}^{2}\rangle=\frac{1}{8}\left[\frac{1}{\omega_{i}}+\frac{\omega_{i}}{|\omega_{f}|^{2}} \right]$ (see Eq.~\eqref{quenches_onemode}). We may go further and attribute this randomness to fluctuations in the time taken to depart from the unstable maximum by defining a delay time $\tau$ through \cite{Guth:1985}
\[
	x=\pm\sqrt{\langle x_{0}^{2}\rangle}e^{|\omega_{f}|\left(t-\tau\right)}.
\]
It is then a straightforward matter to find the distribution of $\tau$
\begin{equation}
	\label{quenches_taudist}
	P(\tau)=\sqrt{\frac{2}{\pi}}|\omega_{f}|e^{-|\omega_{f}|\tau}\exp\left(-\frac{1}{2} e^{-2|\omega_{f}|\tau}\right)
\end{equation}
known as the \emph{Gumbel distribution} in probability theory, where it is often encountered in the statistics of extreme values or -- as here -- in survival analysis.

After the emergence of the classical probability distribution, it is natural that the subsequent evolution be described by deterministic equations. The approach of combining classical evolution with random initial conditions is often known as the `truncated Wigner approximation' \cite{steel:1998}. The effect of finite temperature on the intial state  can be included without difficulty. We will have more to say about the classical field approach in Section~\ref{sub:putting_on_the_brakes}

\subsection{Many modes: quench in the free field theory} 
\label{sub:many_modes_quench_in_the_free_field_theory}

It's now straightforward to generalize the above results to the normal modes of $N$ free quantum fields
\begin{align}
	\label{quenches_modeexpansion}
	\phi_{i}(\br)&=\frac{1}{\sqrt{\text{Vol}}}\sum_{\bk} \sqrt{\frac{1}{2\omega_{\bk}}}e^{i\bk\cdot\br}\left(\adop_{\bk,i}+\aop_{-\bk,i}\right)\nonumber\\
		\pi_{i}(\br)&=\frac{i}{\sqrt{\text{Vol}}}\sum_{\bk} \sqrt{\frac{\omega_{\bk}}{2}}e^{i\bk\cdot\br}	\left(\adop_{\bk,i}-\aop_{-\bk,i}\right),\qquad i=1,\ldots N
\end{align}
The above expressions are for real fields: $\phdop_{i}(\br)=\phop_{i}(\br)$, $\pi_{i}^{\dagger}(\br)=\pi_{i}^{\vphantom{\dagger}}(\br)$. We can form the complex field that appears in Eq.~\eqref{quenches_model} from $N=2$ real fields by taking $\Phi(\br)=\frac{1}{\sqrt{2}}\left[\phi_{1}(\br)+i\phi_{2}(\br)\right]$ and $\Pi(\br)=\frac{1}{\sqrt{2}}\left[\pi_{1}(\br)-i\pi_{2}(\br)\right]$. We will stick with $N=1$ for the remainder of this subsection for clarity: the generalization to many fields is easily written down.

\begin{figure}
	\centering
		\includegraphics[width=0.6\columnwidth]{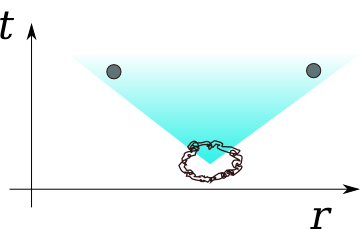}
	\caption{Correlations between two points can only be affected after a quench once they lie in the forward light cone with apex halfway between them at the moment of the quench.}
	\label{fig:light_cone}
\end{figure}

Taking the dispersion relation to be $\omega^{2}_{\bk}=c^{2}\bk^{2}+m^{2}$, we imagine changing the `mass' term  from positive to negative at time $t=0$: $m^{2}_{i}>0\to m_{f}^{2}<0$, leading to imaginary frequencies in the range $|\bk|<|m_{f}|/c$ ($\omega_{\bk}$ appearing in Eq.~\eqref{quenches_modeexpansion} refers to the pre-quench value). We can use the result of Eq.~\eqref{quenches_onemode} to write the second moment $\langle \phi_{\bk}(t)\phi_{-\bk}(t)\rangle$, before Fourier transforming. Slightly neater expressions are obtained by considering the `deep quench' limit in which $m_{i}\gg |m_{f}|$, in which case we make take $\omega_{\bk,i}=m_{i}$ for $|\bk|<|m_{f}|/c$, leading to the integral 
\begin{equation}
	\label{quenches_manymode}
	\langle \phi(\br,t)\phi(\br',t)\rangle \to \frac{m_{i}}{2}\int \frac{d\bk}{(2\pi)^{d}}  \frac{e^{i\bk\cdot\left(\br-\br'\right)}\sinh^{2}\left(t\sqrt{|m_{f}|^{2}-c^{2}\bk^{2}}\right)}{|m_{f}|^{2}-c^{2}\bk^{2}}.
\end{equation}
%
%
The growth of the resulting correlations is determined by causality. To see this, note that for $|\br-\br'|>2ct$ we may close the contour of integration in Eq.~\eqref{quenches_manymode} in the upper half plane of $|\bk|$. There are no poles, so the integral vanishes. Correlations are absent in the initial state by virtue of the large $m_{i}$ limit, and remain zero at time $t$ while the the two points $\br$ and $\br'$ lie outside the forward light cone of a point halfway between them at $t=0$ (see Fig.~\ref{fig:light_cone}). 

Causality places similar (though less severe) restrictions on the propagation of correlations in non-relativistic quantum mechanics, first studied by Lieb and Robinson \cite{lieb1963}.

\subsection{Defect density in the Gaussian approximation} 
\label{sub:defect_density_in_the_gaussian_approximation}								

\begin{figure}
	\centering	\includegraphics[width=0.75\columnwidth]{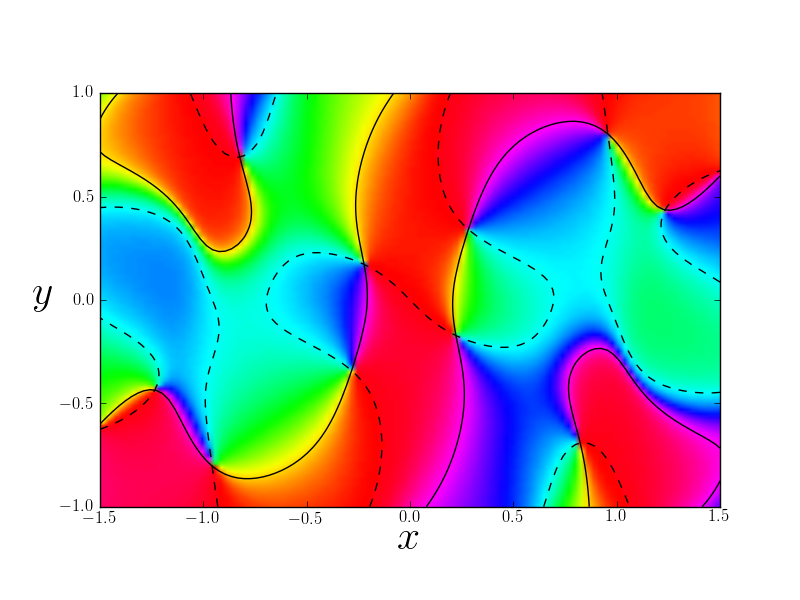}
	\caption{A complex Gaussian random field with correlator given by Eq.~\eqref{quenches_manymode} at $t=2$ (distance is measured in terms of the `Compton wavelength' $c/|m_{f}|$). Solid and dashed lines are the nodes of the real and imaginary parts of the field respectively, with vortices occurring at their intersection. Color indicates the phase of the field.}
	\label{fig:random_field}
\end{figure}

The Gaussian approximation also allows us to address the formation of topological defects. Let us consider the case of vortices in a complex (or two component) field $\Phi(\br,t)$, relevant to the two examples mentioned in Section~\ref{sub:a_model_system}. Usually one thinks of a vortex as an object with a well-defined core region, within which the amplitude of the order parameter goes to zero. Topologically, however, a vortex is a generic zero of a complex field, see Fig.~\ref{fig:random_field}. For Gaussian fields the density of such zeroes must be encoded in the second order correlation function, and is given by the Halperin--Liu--Mazenko formula \cite{halperin:1981,liu:1992}
\begin{equation}
n(t)=-\frac{1}{2\pi}f''(r,t),
\end{equation}
where $f(r,t)\equiv \langle \Phi^{\dagger}(\mathbf{0})\Phi(\br)\rangle/\langle \Phi^{\dagger}(\mathbf{0})\Phi(\mathbf{0})\rangle$ is the normalized correlation function. Applying this result to Eq.~\eqref{quenches_manymode} (or rather its $N=2$ counterpart) gives immediately
\begin{equation}
	\label{quenches_vortexdensity}
	n(t) = \frac{|m_{f}|}{4\pi c^{2} t},\qquad \text{as } t\to\infty.
\end{equation}
It is interesting that even this simple model displays a decay in the density of defects as the order parameter coarsens, leading to vortex-antivortex annihilation\footnote{This is the same behaviour as predicted by the Vinen equation for the decay of the density of vortex lines (see e.g. Ref.~\cite{zurek:1996}) but presumably for quite different reasons.}. Note that interpreting the \emph{quantum} two-point function as a \emph{classical} correlation function hinges upon our earlier discussion of the emergence of the classical picture.

The spontaneous appearance of vortices in the transverse magnetization of a spinor condensate due to this mechanism were observed in the experiment of Ref.~\cite{Sadler2006} that studied quenches in the spin-1 Bose condensate. The expected density of topological defects in this experiment was further discussed in Refs.~\cite{Lamacraft:2007,Uhlmann:2007,Saito:2007}. As mentioned in Section~\ref{sub:a_model_system}, the other appearance of the relativistic `$\lambda \phi^{4}$' in ultracold physics is in the description of the superfluid to Mott insulator transition in the Bose--Hubbard model. Symmetry-breaking quenches from the Mott to superfluid phase were studied in a recent experiment \cite{chen:2011}, though with no conclusive observation of superfluid vortices. A somewhat related observation of vortices formed during the Bose--Einstein condensation transition was reported in Ref.~\cite{Weiler:2008}. 

Taking a step back, we see that the spontaneous formation of topological defects upon passage through a phase transition is intimately connected to the finite propagation time of causal influences. Ordering is destined to proceed independently in regions sufficiently remote from each other, with topological defects a natural consequence. This general picture goes by the name of the \emph{Kibble--Zurek mechanism} (see e.g. Refs.~\cite{kibble1980,zurek:1996}). Ref.~\cite{chuang1991} was an early test of this idea in a liquid crystal.

\subsection{Kibble--Zurek scaling}
\label{sub:kibble_zurek_scaling}

So far we have considered only instantaneous quenches. It is natural to ask what changes when the control parameter is allowed to vary at a finite rate. In this case a simple argument, due to Zurek, describes the overall scale of regions that order independently. Suppose that the control parameter $\delta$ changes linearly with time, $\delta=t/\tau_{Q}$, with the phase transition occurring when $\delta=0$. The correlation length at the transition diverges as $\xi\propto \delta^{-\nu}$, while the characteristic relaxation time of the order parameter $\tau$ is related to $\xi$ by the dynamic critical exponent $z$, and thus displays critical slowing down $\tau\propto \xi^{z}\propto \delta^{-z\nu}$. At some time $t_{*}$ before the transition, the relaxation time exceeds $t_{*}$, so that further growth of the correlation length is impossible. This time scales with $\tau_{Q}$, the quench duration, as $t_{*}\propto \tau_{Q}^{\frac{z\nu}{1+z\nu}}$, corresponding to a length scale $\xi_{*}\propto \tau_{Q}^{\frac{z}{1+z\nu}}$. This characteristic dependence on quench duration is known as \emph{Kibble--Zurek scaling}. An extension of the Gaussian calculations of the previous section to the case of finite quench rate is consistent with this idea \cite{lamacraft2008}.

Considerable theoretical and numerical work has built upon the Kibble-Zurek idea~\cite{Dziarmaga:2005,Zurek:2005,Cherng:2006,silva,pollmannsweep}, but some basic questions remain.  Some derivations rest on the notion of ``excitation''~\cite{polkovnikov:2008} or defect, whose meaning is clear in an integrable model such as the quantum Ising case but not in a generic non-integrable model; for at least some quantities such as entanglement, the behavior after a sweep in a non-integrable model is quite different (it increases more rapidly than in the integrable case).  We are not aware of any experiment showing Kibble--Zurek scaling in an ultracold system. The most convincing tests to date were carried out in superconducting Josephson junctions \cite{monaco2006}.

\subsection{Putting on the brakes} 
\label{sub:putting_on_the_brakes}

The Gaussian approximation used thus far is enlightening but naturally has its limits. Most obviously, the mode amplitudes in the band of unstable wavevectors satisfying $c|\bk|<|m_{f}|$ grow exponentially without limit. In addition, there is no scattering between different momentum states, the usual source of thermalization. In this section we briefly discuss approaches that allow these limitations to be overcome.

It is clear that the inclusion of the quartic term in Eq.~\eqref{quenches_model} will cause the instability to saturate. The Hartree-Fock approximation leads us to replace this term with a quadratic one
\begin{equation}
	\label{quenches_hartree}
	|\Phi|^{4}\to -2\langle|\Phi|^{2}\rangle^{2}+4\langle|\Phi|^{2}\rangle|\Phi|^{2}.
\end{equation}
Note that there are no `anomalous' terms $\langle \Phi^{2}\rangle=0$, reflecting the absence of symmetry breaking in the initial state. The resulting quadratic term can therefore be incorporated into the mass term to given an effective mass
\begin{equation}
	\label{quenches_meff}
	m^{2}_{\text{eff}}=m^{2}+4\lambda\langle|\Phi|^{2}\rangle_{m_{\text{eff}}}.
\end{equation}
The right hand side is $m_{\text{eff}}$ dependent, so this is a self-consistent equation. As in the stationary problem there are ultraviolet divergences to deal with. This method has been used a great deal to study field theories out of equilibrium: some early references are Refs.~\cite{boyanovsky:1993a,cooper:1994}, while Ref.~\cite{Sotiriadis:2010} is a recent study of the quench problem.

The solution of the self-consistent Hartree dynamics saturates the instability, and may be formally justified in the limit of a large number $N$ of field components (for the complex field $\Phi(\br)$ $N=2$) with $\lambda N$ held fixed. But this is cold comfort, for this approximation ignores scattering between the different momentum modes, a \emph{sine qua non} for thermalization.

There are two contrasting pictures of the subsequent evolution. One could take the viewpoint of kinetic theory, using some approximation to close the hierarchy of equations of motion. The derivation of the Boltzmann equation is the most familiar example of this strategy. An enormous literature exists on the corresponding approaches to field theories. Ref.~\cite{berges:2004} provides a pedagogical introduction: it must be admitted that the high energy literature is considerably more sophisticated than the approaches typically encountered in treatments of these problems by condensed matter physicists. The possibility of developing a kinetic description without the `quasiparticle ansatz'\footnote{Essentially this amounts to a separation of scales in which the occupation numbers and spectral function of particles change slowly compared to the typical energy, see e.g. Section 4.2.2 of Ref.~\cite{berges:2004}} could be important for the description of strongly correlated systems out of equilibrium. The second viewpoint focuses on the dynamics of the classical field distribution that emerges after the quench. Usually called the `classical field method', this approach also has a long history, but has enjoyed particular popularity with the advent of Bose condensation in atomic gases, where systems are often dilute and well described by the Gross--Pitaevskii equation. 

Of course these approaches are related. Sticking with the example of the non-relativistic Bose gas (instead of the relativistic case Eq.~\eqref{quenches_model}, where the existence of particle and hole excitations complicates matters), one may derive a kinetic description starting from the classical equations of motion, in which the occupation number corresponds to the mean intensity of the field amplitudes\footnote{The resulting collision integrals only allow for stimulated, and not spontaneous scattering. The latter is negligible for highly occupied modes.}. The very same problem has been studied extensively in the theory of nonlinear waves, where it is known as \emph{wave turbulence} \cite{zakharov:1992}. Closure of the equations is possible for weakly nonlinear, dispersive waves, in which case the amplitudes evolve to a joint Gaussian distribution \cite{benney:1966,Newell:2001}. One expects, however, that the conditions of phase ordering (i.e. accumulation of particles in the low momentum states, in the case of Bose--Einstein condensation) are precisely those under which this approximation breaks down \cite{svistunov:1991,dyachenko:1992}. The introductory section of Ref.~\cite{berloff:2002} provides a useful summary of the difficulties involved in an analytical description of the kinetics of Bose condensation. In this case one would prefer a hybrid scheme in which the long wavelength components of the field (`condensate') are described by a classical equation of motion, with the short wavelength components (`normal fluid') being described by kinetic theory. In such a description the condensate must be subject to noise arising from stochastic interactions with the normal fluid. The derivation of such a `two-fluid' picture also has a long history, see e.g. Refs.~\cite{stoof1999,gardiner2002,griffin:2009,salman:2011,kamenev:2011} for recent treatments. 

Finally, there is the matter of connecting such treatments to the extensive literature studying effective stochastic-dissipative models for equilibrium dynamic critical phenomena (notably the alphabetic classification of Hohenberg and Halperin \cite{hohenberg:1977}\footnote{Ref.~\cite{berges:2010} is a recent example showing that the classical field dynamics of the real scalar `$\lambda \phi^{4}$' theory is consistent with the predictions of Model C, describing a system with non-conserved order parameter coupled to a conserved (energy) density, the latter being a feature of Hamiltonian dynamics.}) and phase ordering \cite{Bray:2002}. 

The two incarnations of the `$\lambda \phi^{4}$' Hamiltonian discussed in Section~\ref{sub:a_model_system} are of recent vintage, but it is puzzling that there has been relatively little fruitful interaction between theory and experiment \footnote{Notable exceptions are Refs.~\cite{kohl:2002,davis:2002,hugbart:2007,ritter:2007,Weiler:2008}.} concerning the kinetics of Bose--Einstein condensation in atomic gases, a phenomenon occurring at the start of almost every experiment in ultracold physics.

\section{Integrability and some consequences}\label{sec:integrability}

As we noted in the introduction, ultracold physics provides a number of situations that come very close to realizing certain integrable models. It is a long-standing conjecture that finite dissipative transport coefficients -- such as the mobility -- can only emerge for systems that are non-integrable \cite{mccoy1994,Castella:1995,saito:1996}. Atomic systems therefore form a natural testbed for this idea, in which one could hope to study transport as integrability is broken to varying degrees. Before continuing we should admit that even giving a precise formulation of the idea of \emph{quantum} integrability is a notoriously thorny matter \cite{caux2010}. To be concrete, we adopt Sutherland's physical point of view that those integrable systems studied in condensed matter (usually by the Bethe ansatz, in one of its avatars) display \emph{non-diffractive scattering} \cite{sutherland2004}. This means that when three or more particles undergo a collision, the only possible outcomes are those in which the set of outgoing momenta coincides with the set of incoming momenta, rather than taking on all possible values allowed by energy and momentum conservation. This statement is evidently equivalent to the existence of a much larger set of conserved quantities of the form $\sum_{i}f(k_{i})$, for some arbitrary function of momentum. It is also equivalent to the factorization of many-body scattering amplitudes into successive two-body scattering events\footnote{For $\delta$-function interactions this is more or less clear.}.

As explained in Section~\ref{sec:introduction}, when this observation is translated into kinetic theory, the resulting lack of relaxation of the distribution function in a 1D integrable system makes the absence of dissipative effects plausible. In this section we discuss this issue with the help of a simple concrete example. 

Let us consider an impurity moving in a spinless Fermi gas. When the impurity and gas masses are equal $M=m$ this problem is integrable for any strength of the interaction $H_{\mathrm{int}}=V\sum_i\delta(x_i-X)$ between the fermions at positions $\{x_i\}$ and the impurity at $X$\footnote{This is a special case of the Gaudin--Yang gas, integrable for any number of `gas' and `impurity' particles \cite{gaudin:1967,yang:1967}.}.

Low energy scattering events near the Fermi surface have momentum transfers close to a multiple of $2p_F$, where $p_F$ is the Fermi momentum. Thus for an impurity of mass $M$ there is a characteristic energy scale $E_{\mathrm{recoil}}\equiv 2p_F^2/M$ associated with a $2p_F$ momentum transfer. At temperatures $k_BT\ll E_{\mathrm{recoil}}$ such processes are frozen out, and the motion of the impurity is determined by forward scattering.

Consideration of the kinematics of these forward scattering events shows that processes involving a single fermion are also suppressed. Let us write the dispersion relation of the fermions as $\xi(p)=p^2/2m-\mu$ for chemical potential $\mu$, and that of the impurity as $\epsilon(p)=p^2/2M$. Scattering of a fermion with momentum $p\sim \pm p_F$ and low momentum transfer $q$ leads to a change in energy of $\xi(p+q)-\xi(p)\sim \pm v_F q$. Since the corresponding change in energy of the impurity is on the order of $q^2/2M$, energy conservation would require $q\sim 2Mv_F=2(M/m)p_F$, or energies $\sim (M/m)^2E_{\mathrm{recoil}}$. This is just a version of the familiar Landau criterion.

\begin{figure}
	\centering
		a)\includegraphics[height=1in]{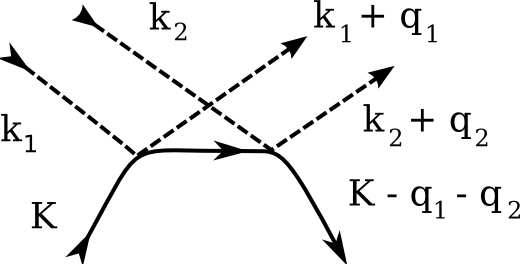}b)\includegraphics[height=1in]{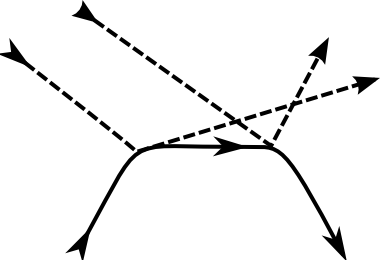}\\
		c)\includegraphics[height=1in]{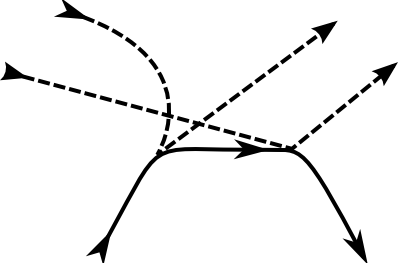}d)\includegraphics[height=1in]{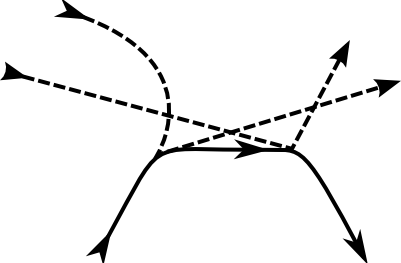}
	\caption{Diagrams contributing to the two-particle scattering amplitude in Eq.~\eqref{2part}. Solid lines denote the impurity, dashed lines the fermions of the gas.}
	\label{fig:2particle}
\end{figure}

These arguments suggest that the low temperature transport of the impurity is due to higher order processes. The simplest such process involves the scattering of two particles with momenta lying close to each fermi point. The lowest order amplitude  for fermions with momenta $k_1$ and $k_2$ to scatter to $k_1+q_1$ and $k_2+q_2$, while the impurity momentum goes from $K$ to $K-q_1-q_2$ (see Fig.~\ref{fig:2particle}) is $\mathcal{T}^{(2)}\delta(E_i-E_f)$ with \cite{lamacraft2008}
%
\begin{equation}\label{2part}
\begin{split}
	\mathcal{T}^{(2)}
	&=i\left(\frac{V}{L}\right)^2\left[\frac{1}{\xi_{k_1+q_1}-\xi_{k_1}+\epsilon_{K-q_1}-\epsilon_K}-\frac{1}{\xi_{k_2+q_2}-\xi_{k_1}+\epsilon_{K-k_2-q_2+k_1}-\epsilon_K}\right.\\
	&\left.+\frac{1}{\xi_{k_2+q_2}-\xi_{k_2}+\epsilon_{K-q_2}-\epsilon_K}-\frac{1}{\xi_{k_1+q_1}-\xi_{k_2}+\epsilon_{K-k_1-q_1+k_2}-\epsilon_K}\right].	
\end{split}
\end{equation}
At low $q_1$, $q_2$ the first and third terms give rise to singular behavior arising forward scattering processes at second order
$\mathcal{T}^{(2)}\to i\left(\frac{V}{L}\right)^2\frac{q_2-q_1}{v_Fq_1q_2}$.
Despite this singularity the momentum relaxation rate of the impurity, given by 
\begin{eqnarray*}
\tau_\mathrm{mom}^{-1}=\frac{2\pi}{\hbar MT}\sum_{k_1,k_2,q_1,q_2}(q_1+q_2)^2|\mathcal{T}^{(2)}|^2\delta(E_i-E_f) n_{k_1}n_{k_2}\left(1-n_{k_1+q_1}\right)\left(1-n_{k_2+q_2}\right)
\end{eqnarray*}
($n_k$ is the Fermi distribution) is finite and vanishes as $T^{4}$, leading to a low temperature mobility $\mu=\tau_\mathrm{mom}/M\propto T^{-4}$~\cite{kagan1986epe,castro-neto1994,castroneto1996}. This calculation, together with the observation that even an almost opaque impurity will appear transparent at low temperatures as the backscattering processes are suppressed, offers a qualitative picture of behavior of the mobility from high to low temperatures. \cite{castroneto1996}

However, the above matrix element in fact \emph{vanishes} in the limit $M=m$ of equal masses, due to interference between the terms. This corresponds to the absence of diffractive scattering mentioned at the beginning of this section, and coincides with the integrable limit. A related observation was made by Gangardt and Kamenev \cite{gangardt:2009} in the case of the Bose gas, where the process analogous to the creation of two soft particle-hole pairs is the creation of two phonons\footnote{To see the same thing happening in simple perturbation theory, one has to consider the integrable point of the bosons plus impurity problem where the boson-boson interaction strength is the same as the boson-impurity interaction, and both the bosons and the impurity have the same mass.}. 

But what about the two-particle scattering processes? Though they may be exponentially suppressed at low energies, they can certainly change the momentum of the impurity. It is not immediately clear from this picture why ballistic motion of the impurity is guaranteed when $M=m$. To understand why this is so, we use a variant of the Kohn formula for the Drude weight \cite{kohn1964,Castella:1995} to express the singular part of the mobility $\text{Re}\,\mu(\omega)=\frac{\pi}{M} D\delta(\omega)+\cdots$ as
\begin{equation}
	\label{quenches_Kohn}
	D = ML^{2}\sum_{n} p_{n} \frac{\partial^{2}E_{n}}{\partial \Phi^{2}}
\end{equation}
where the sum is over the many-body eigenenergies $E_{n}$ and $p_{n}$ are the corresponding Boltzmann weights. $\Phi$ is a flux felt only by the impurity particle. For a free particle $D=1$. Now we invoke another feature of integrable systems: the occurrence of level crossings in the spectrum with the variation of a parameter that preserves integrability, of which $\Phi$ is an example. Because of the absence of avoided crossings, it is plausible that the average curvature is nonzero in the thermodynamic limit. When $M\neq m$, avoided crossings occur and $D=0$ may result. Ref.~\cite{Castella:1995} studied a lattice version of this model via exact diagonalization and the Bethe ansatz in the integrable case, finding some support for this conjecture.

We are left to ponder the connection between nonzero $D$ and non-diffractive scattering. Some insight is gained by considering the curvature of the free energy with respect to the flux \cite{zotos:1997}
\begin{equation}
	\label{quenches_FreeCurv}
\begin{split}
	\frac{\partial^{2} F}{\partial \Phi^{2}}&=\frac{D}{ML^{2}}-\frac{\beta}{(ML)^{2}}\left(\sum_{n}p_{n}\langle n|P|n\rangle^{2}-\langle P \rangle^{2}\right)\\
	&=\frac{D}{ML^{2}}-\frac{\beta}{(ML)^{2}}\lim_{T\to\infty}\frac{1}{T} \int_{0}^{T}dt(\langle P(t)P(0)\rangle - \langle P\rangle^{2})
\end{split}	
\end{equation}
$\partial^{2} F/\partial \Phi^{2}=0$ in the thermodynamic limit, so
\begin{equation}
	\label{quenches_Dbound}
	D =\frac{\beta}{M} \lim_{t\to\infty}(\langle P(t)P(0)\rangle - \langle P\rangle^{2}).
\end{equation}
Thus a non-zero $D$ corresponds to non-decaying momentum correlations. By Khinchin's theorem \cite{khinchin:1949}, this amounts to a failure of ergodicity for the variable $P$. The relation between this behavior and the conservation laws is provided by the \emph{Mazur inequality} \cite{mazur:1969}
\begin{equation}
	\label{quenches_MazurIneq}
	\lim_{t\to\infty}(\langle P(t)P(0)\rangle - \langle P\rangle^{2}) \geq \sum_{\alpha}\frac{\langle P Q_{\alpha}\rangle^{2}}{\langle Q_{\alpha}Q_{\alpha}\rangle},
\end{equation}
where $Q_{\alpha}$ are the conserved quantities, orthogonal to each other under the inner product provided by $\langle \cdots\rangle$. Eq.~\eqref{quenches_MazurIneq} is in fact an equality if \emph{all} conserved quantities are included \cite{suzuki:1971}. The resulting connection between ballistic transport, the (thermally averaged) level curvature, and the presence of conserved quantities has been explored in a number of works: the introductory sections of Ref.~\cite{sirker:2010} provide a detailed summary.

\begin{figure}
\centering
	\includegraphics[height=3in]{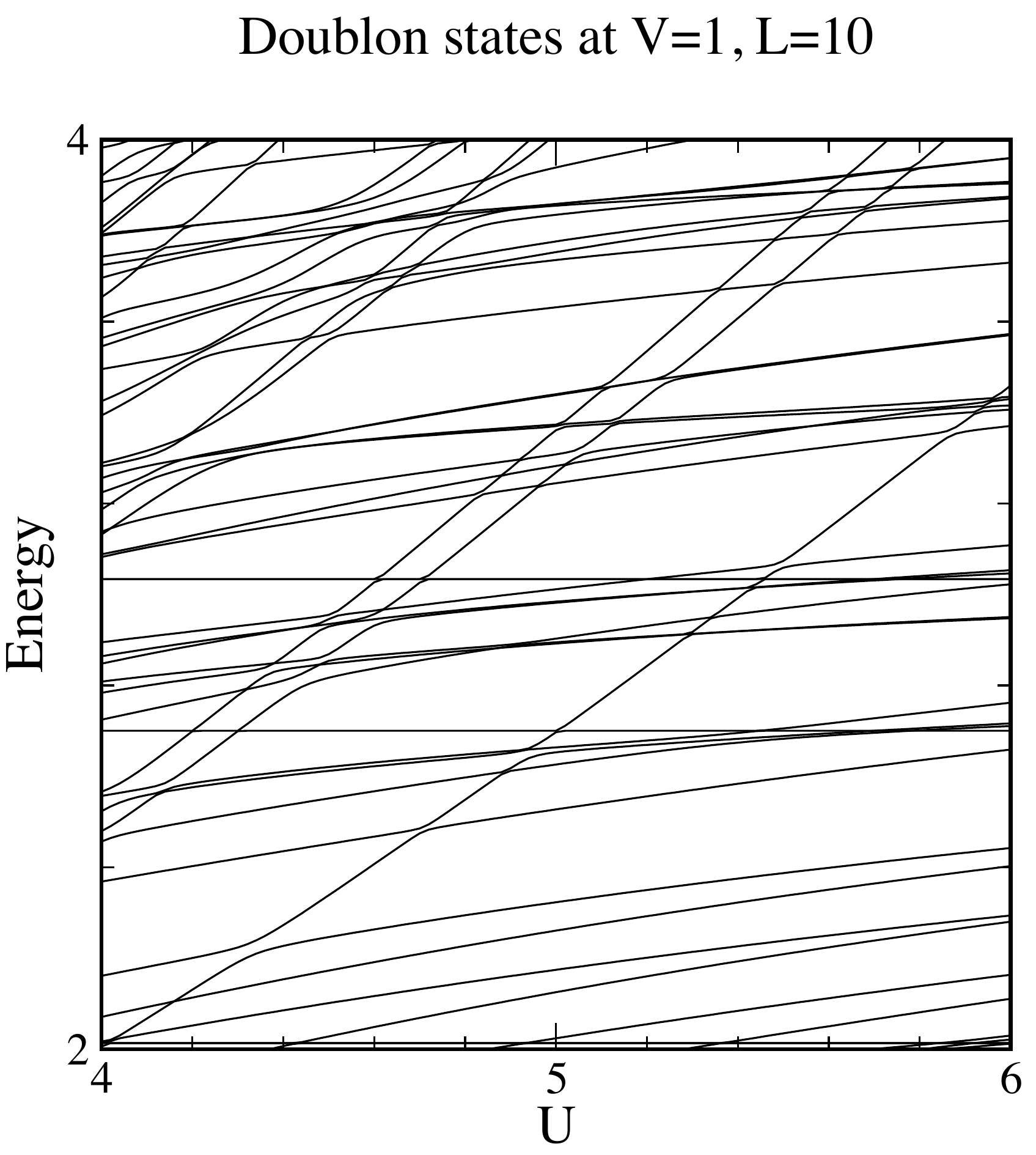}
\caption{Spectral signature of doublon decay in the 1D extended Hubbard model, with on-site interaction $U/t$, and nearest neighbor interaction $V/t$ breaking the integrability of the model. The steep diagonal lines correspond to states of the system with a single doublon; the shallow lines to states with no doublon. The mixing between the two types of states that results in avoided crossings is the spectral signature of doublon decay that would be absent at $V=0$.}
\label{fig:DoublonDecay_Closeup}	
\end{figure}

We close this section by pointing out that the above ideas are of more general applicability. Consider the example of \emph{doublon decay}, the decay of doubly occupied states in the fermion Hubbard model, already observed in ultracold gases \cite{Strohmaier:2010}. In the 1D Hubbard model
\begin{equation}
	\label{quenches_1DHubb}
	H_{\text{Hubbard}}=-t\sum_{s,i}\left[\cdop_{s,i}\cop_{s,i+1}+\cdop_{s,i+1}\cop_{s,i}\right]+U\sum_{i}n_{\uparrow,i}n_{\downarrow,i},
\end{equation}
this process is forbidden by integrability, but it can occur when integrability is broken. The doublon number operator is $N_{d}\equiv\sum_{i} n_{\uparrow,i}n_{\downarrow,i}$, where $n_{s,i}$ denotes the number of spin $s$ particles on site $i$. But since this is just the on-site interaction term the above reasoning implies
\begin{equation}
	\label{quenches_DoublonDecay}
	\sum_{n} p_{n}\frac{\partial^{2} E_{n}}{\partial U^{2}} < \beta \lim_{t\to\infty}(\langle N_{d}(t)N_{d}(0)\rangle - \langle N_{d}\rangle^{2})
\end{equation}
where we now have an inequality because the `susceptibility' $-\partial^{2} F/\partial U^{2}$ should be positive for stability. Doublon decay is thus associated with a vanishing of the averaged level curvature with respect to $U$, see Fig.~\ref{fig:DoublonDecay_Closeup}.

\section{Disordered systems and many-body localization} \label{sec:loc}  





A fertile area of condensed matter research has been the physics of systems with ``quenched'' disorder: there are random quantities in the Hamiltonian that are not included in the thermal average.  One may be interested in either the mean values of physical quantities after some non-thermal average over the disorder realizations, or in typical values that are the most likely to be observed even if rare realizations dominate the average.  The possibility of retaining quantum coherence for long times in ultracold atomic systems may allow experimental answers to some basic questions in the theory of localization in quantum disordered systems.  The `Anderson localization' of a single particle moving quantum-mechanically in a random potential is in many cases well understood.  Remarkably, a particle moving in one or two spatial dimensions has only localized energy eigenstates (i.e., the wavefunction falls off exponentially at spatial infinity) even for a weakly disordered potential; this is in contrast to the case of three dimensions, where a ``mobility edge'' separates localized wavefunctions at low energies from extended wavefunctions at high energies.

The effect of interparticle interactions on this description has been a subject of active research for many years but is still not completely understood~\cite{lee:1985}.  For experiments on electrons in solids, the standard approach for describing experiments is to assume that an electron loses its phase coherence after some temperature-dependent dephasing time.  This dephasing time provides a cutoff on the localization physics, as localization is a consequence of phase coherence.  Perturbative treatments of interparticle interactions have had success in describing some experiments, but at least one basic question remains unanswered.

The dephasing time is an example of interaction with a ``bath'', similar to those discussed earlier: the electron system in a disordered metal is treated as open rather than Hamiltonian.  It is natural to ask what happens in an interacting particle system at nonzero temperature when there is no external bath.  (Since there is no bath, the meaning of ``temperature'' is simply a weight on initial conditions for Hamiltonian dynamics, as in the Kubo formula.)  From our discussion of equilibration, one might think that one part of the system can see the rest of the system as such a bath, at least approximately, so that interactions alone lead to an apparent dephasing time.

Recent theoretical work suggests that there can be a localization transition in a closed (i.e., Hamiltonian) interacting system at some finite temperature 
$T_c$, when all single-particle states are localized~\cite{basko:2006}. Above $T_c$ the system acts as its own bath and appears dephased on long length or time scales, while below $T_c$ the localization persists.  Actually the problem can be simplified somewhat by switching from continuum `particle' systems with an unbounded entropy per unit volume at infinite temperature to lattice systems with finite maximum entropy per site~\cite{Oganesyan:2007}.  Then the many-body localization transition exists as a function of parameters even at infinite temperature, and we will discuss an example believed to exhibit such a transition.

We start from an $XX$ spin chain with random Zeeman magnetic fields:
\begin{equation}\label{chain}
H_0 = - J \sum_i \left( S_i^x S_{i+1}^x + S_i^y S_{i+1}^y \right) - \sum_i h_i S_i^z.
\end{equation}
Here the fields $h_i$ are drawn independently from the interval $[-\Delta,\Delta]$.  The eigenstates of this model are equivalent through the Jordan--Wigner transformation to Slater determinants of free fermions with nearest-neighbor hopping and random on-site potentials.  Since every single-fermion eigenstate is localized by those potentials as long as $\Delta > 0$, we expect that the dynamics of this spin Hamiltonian are localized as well.  The response of the system to a local perturbation is confined for all times with exponential accuracy to a radius set by the single-particle localization length.

Now if instead we had started with Heisenberg rather than $XX$ couplings between the spins, i.e., added a term $- J \sum_i S_i^z S_{i+1}^z$, then the model, at infinite temperature, is thought to have a dynamical transition as a function of the dimensionless disorder strength $\Delta/J$~\cite{Oganesyan:2007,Pal:2010}.  This transition should be manifest in several physical quantities, including level statistics and dynamical correlation functions.  Level statistics should transition from Wigner-Dyson statistics and level repulsion for small $\Delta/J$ to the Poissonian distribution characteristic of a disordered system (since spatial regions separated by the localization length are effectively independent) for large $\Delta/J$.  Correlation functions of the conserved quantity $S_z$ (equivalent to particle number) should show, as originally emphasized by Anderson~\cite{anderson:1958}, a transition in the long-time behavior between particles escaping to infinity or remaining in a bounded region.

A system in the localized phase may fail to come into mutual thermal equilibrium among its constituent pieces. To understand why this is so, consider dividing the above spin chain into two equal halves, A and B. Though the disorder is not the same on each side, we expect the two halves to be macroscopically identical. Conventional thermodynamic wisdom would have that in the long time limit, the expectation value of total spin $\mathcal{S}^{z}_{A/B}=\sum_{i\in A/B}S_{i}^{z}$ is the same on both sides. 

Let us take the initial state of the system to be
\begin{equation}
	\label{atommanybodyloc_init}
	\ket{\Psi}=\ket{M_{A}}_{A}\otimes\ket{M_{\text{tot}}-M_{A}}_{B}
\end{equation}
where $\ket{M_{A/B}}_{A/B}$ are states of the A and B subsystems with $\mathcal{S}^{z}_{A/B}=M_{A/B}$. Evidently there are many such states. In terms of the eigenstates of the whole system, the long time average of $\Delta \mathcal{S}^{z}=\mathcal{S}^{z}_{A}-\mathcal{S}^{z}_{B}$ is then 
\begin{equation}
	\label{atommanybodyloc_longtime}
	\overline{\langle\Delta \mathcal{S}^{z} \rangle}=\sum_n \langle n|\Delta \mathcal{S}^{z}|n\rangle |\braket{n|\Psi}|^{2}.
\end{equation}
In this sum the first factor reflects the distribution of $\Delta \mathcal{S}^{z}$ in the eigenstate $\ket{n}$, while the second factor reflects the overlap of this state with the initial state. 

Let us consider the uncoupled system, where the appropriate hopping term in Eq.~\eqref{chain} that connects the two regions is set to zero. The eigenstates $\ket{n}$ are then eigenstates of $\Delta \mathcal{S}^{z}$, and no relaxation occurs. The vast majority of states have $\Delta \mathcal{S}^{z}$ eigenvalue $\Delta M\sim 0$, and the distribution of $\Delta M$ generally has the form $P_{s}(\Delta M)=\exp\left[-N s(\Delta M/N)\right]$: deviations from the equilibrium value are exponentially rare. The $s$ subscript is to remind us that the function $s(x)$ is just the entropy (per site).

When a small coupling $J$ is switched on, states with different $\Delta M$ are mixed together, but we expect them to retain their character, meaning that the histogram of $\Delta M$ probabilities associated with a particular state will be strongly peaked around the value that it had with certainty for $J=0$. Similarly, the probabilities $|\braket{n|\Psi}|^{2} $ will be largest for those states peaked around $\Delta M_{0}\equiv 2M_{A}-M_{\text{tot}}$, the initial value. Let us group the eigenstates in sets $\mathcal{N}_{\Delta M}$ labelled by the value of $\Delta M$ that they had when $J=0$. Then defining, $P_{\Psi}(\Delta M)\equiv \sum_{n\in \mathcal{N}_{\Delta M}}|\braket{n|\Psi}|^{2}$, Eq.~\eqref{atommanybodyloc_longtime} has the form
\begin{equation}
	\label{atommanybodyloc_longcoarse}
	\overline{\langle\Delta \mathcal{S}^{z} \rangle}\sim\sum_{\Delta M} \Delta M P_{s}(\Delta M)P_{\Psi}(\Delta M).
\end{equation}
Since states in $N_{\Delta M}$ only overlap with $\ket{\Psi}$ at the $|\Delta M-\Delta M_{0}|^{\text{th}}$ order of perturbation theory in $J$, it is reasonable that $P_{\Psi}(\Delta M)$ is exponentially small is this deviation. Thus Eq.~\eqref{atommanybodyloc_longcoarse} is seen to involve a competition between two exponential factors that in the thermodynamic limit can pick out a value of $\overline{\langle\Delta \mathcal{S}^{z} \rangle}$ different from zero, contradicting the assumption that the two parts of the system come to mutual equilibrium\footnote{The importance of rare states in the absence of thermalization was emphasized in Ref.~\cite{biroli:2010}}.

The crucial part of the picture is that the states `retain their character' at $J\neq 0$, and this is just what is implied by the recent work discussed above. When there is no longitudinal part to the interaction, this is just Anderson localization of noninteracting Jordan--Wigner fermions. What is new is the claim that the same remains true in the interacting problem over some range of $J$.

However, the location of the transition and its critical properties have been difficult to obtain even numerically.  A search based on level statistics was inconclusive~\cite{Oganesyan:2007}, while studying correlation functions suggests a phase transition~\cite{Pal:2010}  In both cases exact diagonalization of the Hamiltonian was used, which  strictly limits accessible system sizes.  The phase transition in the correlation functions was interpreted in terms of an ``infinite-disorder'' critical point, analogous to the critical points in ground-state quantum phase transitions in disordered systems accessible via the real-space renormalization group (RSRG)~\cite{monthusreview,refaelreview}.   A approach to many-body localization based on numerical solution of RSRG equations predicts a sharp transition and a specific location~\cite{Monthus:2010}, but it is currently difficult to compare these predictions to numerics on the microscopic model.

Specific proposals for realizing a many-body localization transition with ultracold atoms are discussed in Ref.~\cite{Aleiner:2009}.  Another way of searching for the putative many-body localization transition, at least numerically, involves the entanglement between different spatial regions.  In the disordered phase, one would naively expect the entanglement entropy, defined in a moment, to saturate, as indeed happens in the non-interacting case, while in the extended phase entanglement would spread throughout the system.  We discuss an interesting behavior in the dynamics of entanglement in the disordered phase below, after explaining some general features of entanglement in many-body systems.

\section{Entanglement dynamics}

One of the most significant potential advantages of interacting many-atom systems is the ability to maintain quantum coherence for relatively long times than in the electron subsystem of solids.  Considerable effort has been devoted in recent years to understanding how entanglement, one of the basic notions of quantum information, behaves in many-particle systems, both for intrinsic interest and for potential application to quantum computing.  Since atomic systems at the moment may offer the best hope for observing many-particle entanglement in a quantitative way, and since entanglement underlies many of the questions touched on in preceding sections, we now review the basic notions of entanglement and its dynamics in model atomic Hamiltonians.

\subsection{Basics of entanglement in ground states of many-body systems}

In our discussion of entanglement, for simplicity we will limit ourselves to the case of a pure state of a bipartite system $AB$.  While the Hilbert space of the full system is spanned by product states $|\psi_i \rangle \otimes |\phi_j \rangle$ of basis states of $A$ and $B$, there are superpositions of the basis states that cannot be factored into {\it any} pure states of $A$ and $B$.  A familiar example is the singlet state of two spins: $|\Psi_{AB} \rangle = \frac{1}{\sqrt{2}} \left( |\uparrow_A \rangle \otimes |\downarrow_B \rangle - |\uparrow_B \rangle \otimes |\downarrow_A \rangle \right).$  An entangled state is simply one that is not a product state.   The entanglement entropy, which has been the quantity most studied in the context of many-particle systems, is defined as the von Neumann entropy of the reduced density matrix for either subsystem,
\begin{equation}
S = -{\rm Tr} \rho_A \log \rho_A = -{\rm Tr} \rho_B \log \rho_B,
\end{equation}
where $\rho_A$ ($\rho_B$) is the reduced density matrix of subsystem $A$($B$) obtained by tracing over degrees of freedom in the other subsystem.

The basic behavior of entanglement entropy in ground states of standard atomic or condensed matter is generally well understood, especially in one dimension.   We will not discuss ground-state entanglement entropy in detail here but simply state a few results before referring the reader to recent in-depth reviews.  In gapped systems, the general expectation, which can be proved in some cases, is that entanglement entropy satisfies an ``area law'', i.e., scales as the volume of the $AB$ boundary.  For regular geometries, the scaling is thus as $L^{d-1}$ for a subregion $A$ of linear size $L$ cut out of an infinite $d$-dimensional system.  Topological phases can have subleading terms in the entanglement entropy that are of order unity (i.e., scaling as $L^0$), in addition to the area law, and probe aspects of the topological order~\cite{Levin05b,Kitaev06b}.

For critical systems, the behavior of ground-state entanglement entropy is more complicated.  In one dimension, for quantum critical points described by 2D conformal field theories, the entanglement entropy diverges logarithmically, with a universal coefficient determined by the central charge~\cite{holzhey,vidalent,Calabrese05}:
\begin{equation}
S \approx {c \over 3} \log L/a \quad {\rm as}\ L \rightarrow \infty
\end{equation}
for a block of length $L$ cut from an infinite chain, where $c$ is the central charge and $a$ a short distance cutoff.
There are similar logarithmic divergences at infinite-randomness critical points in one dimension~\cite{refaelreview}, although differences between the pure and random cases~\cite{santachiara} become manifest in the entanglement spectrum~\cite{fagotti}.  In higher dimensions, critical points with a Fermi surface can violate the area law~\cite{wolf,klich}, while others can obey the area law but also have subleading corrections of interest~\cite{fradkinreview}.

To date it has been difficult to observe entanglement directly except in small systems~\cite{monroedensity}.  Since many-particle systems with local Hamiltonians are expected to have entanglement described by the area law, or possibly larger by a logarithmic factor, thermal entropy (which scales as volume $L^d$ under the circumstances above) will rapidly dominate except at the lowest temperatures.  However, there are recent proposals~\cite{klichlevitov,cardyentanglement} for how relatively standard measurement processes could probe many-particle entanglement.  Entanglement also determines the ability of certain numerical methods based on matrix product states and their generalizations to capture accurately the ground state of a given Hamiltonian.

\subsection{Entanglement dynamics of low-lying states}

The growth of entanglement after a quench or sweep of parameters in the Hamiltonian has been studied for one-dimensional systems using similar methods to those studied above in Sections~\ref{sec:quench}.  For a quench in a conformally invariant system, one can distinguish two simple scenarios: a ``global quench'' in which the parameters are changed instantly everywhere in space, versus a ``local quench'' in which the parameters are only changed at one point~\cite{Calabrese:2006}.  In both cases, entanglement propagates along ``light cones''; this can be understood intuitively by noting that conformal invariance implies that all excitations propagate with a single velocity.  For a global quench, this leads to entanglement growing linearly in time after the initial 

The same argument can be applied to understand the effects of sweeping across a conformally invariant critical point.  We focus on the case of a translation-invariant problem, which is similar to the global quench as far as how entanglement propagates: after the sweep, if the Hamiltonian is held constant then entanglement entropy increases linearly in time.  The rate of this linear increase, however, is now determined in simple models by the number of excitations that were created above the ground state when the energy gap was small (we make the same assumptions here as in Section~\ref{sub:kibble_zurek_scaling}).  This number is determined as a function of the sweep rate $\Gamma$ by the same scaling arguments as presented above~\cite{polkovnikov}.  Right after the sweep, the entanglement scales as a logarithm of the sweep rate with a universal coefficient combining central charge $c$ and correlation length exponent $\nu$~\cite{Cincio:2007p052321,pollmannsweep}; for the half-chain entropy,
\begin{equation}
S = -\frac{c\nu}{6(\nu+1)} \log \Gamma+\text{const},
\label{Eq:scalent}
\end{equation}

In more complicated models with interactions between the ``excitations'' in the final gapped region of the phase diagram, the behavior of entanglement during a sweep is more complex.  Without conformal invariance, there is no general rule known for the growth of entanglement with time, and indeed entanglement seems to grow quite rapidly once excitations begin to interact~\cite{pollmannsweep}.  (Numerical studies in this regime become technically challenging as the entanglement quickly saturates the amount that can be captured with matrix product states of a given matrix size.)

The growth of entanglement can be understood as a consequence of the ``entanglement thermalization hypothesis''~\cite{Deutsch:1991,Rigol:2008}: if the interactions lead to apparent equilibration for local measurements with an effective temperature determined by the (constant) energy density, then the entanglement entropy of a subregion must reproduce the thermal entropy at that temperature, since this entropy can in principle be measured.  But the thermal entropy will satisfy a volume law rather than an area law, and hence will be parametrically larger for a large subregion than the initial entanglement entropy, as the system began close to a ground state satisfying the area law (possibly with logarithmic corrections).

In closing, we note that the putative many-body localized phase discussed above also has a signature in the dynamics of entanglement that is not yet well understood.  In a localized phase of a non-interacting system, an initial product state has entanglement that grows in time but saturates once the individual particles have diffused over a length scale set by their localization length.  In an extended phase, one expects entanglement to increase more rapidly, presumably as some power law of time depending on whether the transport is diffusive or ballistic.  In interacting disordered systems, numerics suggest~\cite{prelovsek,manybodylocunpub} a regime where the entanglement entropy increases {\it logarithmically} in time without limit.  Whether this is generic behavior and indicates that the phase is not truly localized, or instead is a qualitative difference between localized phases with and without interactions, are questions for future theoretical work.



\section{Acknowledgements}
\label{sec:acknowledgements}

A.L. acknowledges the support of the NSF through grant DMR-0846788 and Research Corporation through a Cottrell Scholar award, and useful discussions with Peter Arnold, Alex Kamenev, Cass Sackett, and Dan Stamper-Kurn. J.E.M. was supported by the ARO OLE program.


\end{document}